\documentclass{aa}
\usepackage{graphicx}
\usepackage{txfonts}
\def\swift{{\em Swift\/}}
\def\sax{{\em Beppo}SAX\/}

\begin{document}
   \title{Swift XRT Observations of the Breaking X--ray Afterglow \\
          of GRB 050318}


   \author{M. Perri\inst{1}, P. Giommi\inst{1,2}, M. Capalbi\inst{1},
           L. Amati\inst{3}, F. Frontera\inst{3,4}, G. Chincarini\inst{5,6}, S. Campana\inst{5},
           A. Moretti\inst{5}, \\ P. Romano\inst{5},
           G. Tagliaferri\inst{5}, G. Cusumano\inst{7}, V. La Parola\inst{7}, 
           V. Mangano\inst{7}, T. Mineo\inst{7}, D.N. Burrows\inst{8}, J.E. Hill\inst{9,10}, \\ 
           J.A. Kennea\inst{8}, D.C. Morris\inst{8}, C. Pagani\inst{5,8}, J.A. Nousek\inst{8}, 
	   L. Angelini\inst{9}, N. Gehrels\inst{9}, M. Still\inst{9}, A.P. Beardmore\inst{11}, \\ 
           M.R. Goad\inst{11}, O. Godet\inst{11}, P.T. O'Brien\inst{11}, J.P. Osborne\inst{11}, 
           K.L. Page\inst{11}
          }
          
   \offprints{\email{perri@asdc.asi.it}}

   \institute{ASI Science Data Center, Via Galileo Galilei, 
              I-00044 Frascati, Italy
	 \and Agenzia Spaziale Italiana, Unit\`a Osservazione dell'Universo,
              Viale Liegi 26, I-00198 Roma, Italy
	 \and INAF -- Istituto di Astrofisica Spaziale e Fisica Cosmica,
              Sezione di Bologna, Via Gobetti 101, I-40129 Bologna, Italy
	 \and Universit\`a degli Studi di Ferrara, Dipartimento 
              di Fisica, Via Saragat 1, I-44100 Ferrara, Italy
	 \and INAF -- Osservatorio Astronomico di Brera, 
              Via Bianchi 46, I-23807 Merate, Italy
	 \and Universit\`a degli Studi di Milano-Bicocca, Dipartimento 
              di Fisica, Piazza delle Scienze 3, I-20126 Milano, Italy
	 \and INAF -- Istituto di Astrofisica Spaziale e Fisica Cosmica,
              Sezione di Palermo, Via La Malfa 153, I-90146 Palermo, Italy
	 \and Department of Astronomy \& Astrophysics, Pennsylvania State 
              University, University Park, PA 16802, USA
	 \and NASA/Goddard Space Flight Center, Greenbelt, MD 20771, USA
	 \and Universities Space Research Association, 10211 Wincopin Circle, Suite 500, 
              Columbia, MD, 21044-3432, USA
	 \and Department of Physics \& Astronomy, University of Leicester,
              Leicester LE1 7RH, UK
             }

   \authorrunning{M. Perri et al.}
   \titlerunning{Swift XRT Observations of the Breaking X--ray Afterglow of 
                 GRB 050318}
   \date{Received / Accepted}

   \abstract{We report the results of \swift\ X--Ray Telescope (XRT) observations of 
GRB 050318. This event triggered the Burst Alert Telescope (BAT) aboard 
\swift\ and was followed-up with XRT and UVOT for 11 consecutive orbits starting 
from 54 minutes after the trigger. A previously unknown fading X--ray source was detected 
and accurately monitored. The source was found to decrease in intensity with time and a clear 
temporal break occurring at $\sim$18000 s after the trigger was observed. The X--ray light curve was 
found to be consistent with a broken power-law with decay indices $-1.17\pm 0.08$ and 
$-2.10^{+0.22}_{-0.24}$ before and after the break. The spectrum of the X--ray afterglow 
was well described by a photoelectrically absorbed power-law with energy index of $-1.09\pm0.09$. 
No evidence of spectral evolution was found. We compare these results with those obtained with
UVOT and separately reported and refine the data analysis of BAT. We discuss our
results in the framework of a collimated fireball model and a synchrotron radiation
emission mechanism. Assuming the GRB redshift derived from the farthest optical absorption
complex ($z = 1.44$), the event is fully consistent with the $E_{\rm{p}}$--$E_{\rm{iso}}$ correlation.

   }

  \maketitle

\section{Introduction}
\label{intro}
\indent

Following the crucial step forward in the knowledge of the Gamma--Ray Burst (GRB) phenomenon
accomplished with the {\it Beppo}SAX satellite, great expectations are placed in the
\swift\ mission. The \sax\ discovery of GRB afterglows with the resulting determination 
of the distance scale of long events (duration greater than 2 seconds), allowed, among other things, 
the first spectral and temporal studies of the late (i.e. after 0.25 days) X--ray afterglow 
(see, e.g., the review by \cite{Frontera03}). 
However, no observation of the early afterglow, except in one case (GRB 990123,
Maiorano et al. 2005\nocite{Maiorano05}), was possible with {\em Beppo}SAX\/. 
Indications about an afterglow onset during the tail of the prompt emission could be inferred 
only from the comparison of the prompt with the delayed X--ray emission (\cite{Frontera00}). 
No evidence of temporal breaks in the late afterglow light curves 
could be detected, although the X--ray data were found to be consistent with the afterglow 
breaks observed in the optical band (see, e.g., \cite{Pian01}, \cite{Zand01}).
The true onset of GRB X--ray afterglows, their light curves at early times, their spectra, 
the existence of time breaks in the X--ray afterglow light curves are among the issues 
left open by \sax\ and still unsolved despite the availability of large X--ray observatories, 
like {\it Chandra} or {\it XMM--Newton}.

The \swift\ Gamma--ray Burst Explorer (\cite{gehrels}) was successfully launched 
on 2004 November 20. The scientific payload of the observatory consists 
of a wide-field instrument, the gamma--ray (15--350 keV) Burst Alert Telescope 
(BAT, \cite{barthelmy}) which detects bursts and determines their 
position with $\sim$3 arcmin precision, and two co-aligned narrow-field 
instruments: the X--Ray Telescope (XRT, \cite{burrows05a}), operating in the 
0.2--10 keV energy band with source localization capabilities of 5--6 arcsec, 
and the Ultraviolet/Optical Telescope (UVOT, \cite{roming}), sensitive 
in the 170--600 nm band with $\sim$0.3 arcsec positioning accuracy. 

The autonomous and rapid slewing capabilities of \swift\ allow the 
prompt (1--2 minutes) observation of GRB afterglows with the XRT and 
UVOT instruments. 
In particular in the X--ray energy band, where the reaction times of other 
satellites are limited to time scales of several hours, it is possible 
to study the afterglows in their previously unexplored early phases.
At the time of writing (June 2005), the XRT has performed follow-up 
observations of 30 bursts discovered by the BAT instrument and detected in 
all cases their X--ray afterglows (e.g. \cite{burrows05b}, \cite{Campana05}, 
\cite{Tagliaferri05}, \cite{cusumano}, Page et al.~2005).

In this letter we present a detailed analysis of the XRT follow-up observation of 
GRB 050318. The analysis of UVOT data has been presented in a separate paper 
(\cite{Still05}). 
All the uncertainties on the derived quantities are given
at 90\% confidence level for one interesting parameter. Temporal and spectral indices are written 
following the notation $F(t,\nu) \propto t^\alpha \nu^\beta$.

\section{Observations and BAT results}
\label{bat}
\indent

The BAT discovered and located this GRB at 15:44:37 UT on 2005 
March 18 (\cite{krimm05a}). The BAT position (RA(J2000)=49\fdg651, Dec(J2000)=$-$46\fdg392 
with a 90\% containment radius of 3 arcmin) was distributed via the 
GRB Circular Network (GCN). This position was within the \swift\ Earth 
horizon constraint and the spacecraft had to execute a delayed automatic slew 
to the burst position, settling on target at 16:39:11 UT, when XRT and UVOT 
observations started.
The XRT detection of the X--ray afterglow was reported shortly afterwards together with a more 
accurate (6 arcsec) X--ray position determination (\cite{nousek}, \cite{beardmore}).
The afterglow detection by UVOT in the $U$, $B$, and $V$ bands was also 
soon distributed via GCN (\cite{mcgowan}, \cite{depasquale}).
Ground-based follow-up observations in the optical band with the 6.5m 
Magellan/Baade telescope (\cite{mulchaey}) also led to the detection of 
two absorption systems at \mbox{$z=1.20$} and \mbox{$z=1.44$}, indicating
as likely redshift of the event \mbox{$z=1.44$} (\cite{berger}). 

In the BAT prompt emission light curve three peaks are observed. The first peak is followed by a 
quiet phase lasting $\sim$17 seconds and then by two overlapping peaks. 
On the basis of the preliminary analysis (\cite{krimm05b}), the burst is characterized 
by a duration of $T_{90} = 32\pm2~\mathrm{s}$, a fluence in the 15--350 keV band of 
$2.1\times10^{-6}~\mathrm{erg}~\mathrm{cm}^{-2}$ and a time averaged 
15--350 keV spectrum described by a power--law model with energy index 
$\beta = -1.1\pm0.1$ at 90\% confidence level.
We refined the preliminary spectral analysis by grouping the data  points in order 
to increase the significance of each
spectral bin at energies above $\sim$100 keV, where the signal is weak. 
We find that, in the 18--200 keV energy band, a cut--off power--law 
($F(E)\propto E^{\beta}\exp{(-E/E_0)}$)
fits the BAT spectrum significantly better ($\chi_r^2=0.83$ with 8 degrees of freedom, dof) 
than a simple power--law ($\chi_r^2 = 2.04$, 9 dof), with an 
energy index $\beta = -0.34_{-0.33}^{+0.31}$ and $E_0 = 71 _{-26}^{+75}$~keV.
The corresponding peak energy $E_{\rm{p}}$ of the $\nu F(\nu)$ spectrum is 
$E_{\rm{p}} = (1+\beta)E_0 = 47_{-8}^{+15}$~keV, in the uncertainty of which the 
covariance between  $E_0$ and $\beta$ is taken into account. 
From these results and assuming as redshift $z = 1.44$, the rest frame peak 
energy is $E_{\rm{p,rest}} = 115_{-20}^{+37}$~keV, while the isotropic-equivalent 
radiated energy in the rest frame 1--$10^4$ keV energy band is 
$E_{\rm{iso}} = (2.20\pm 0.16) \times 10^{52}$~erg determined with 
the same method adopted by  Amati et al.~(2002)\nocite{Amati02} 
and Ghirlanda et al.~(2004)\nocite{Ghirlanda04} with the following  
cosmological parameters: $\Omega_{\rm{m}} = 0.3$, $\Omega_{\rm{\lambda}} = 0.7$ and
$H_0 = 70$~km~s$^{-1}$~Mpc$^{-1}$. 
Results on the spectral evolution of the prompt emission are reported
by Still et al. (2005)\nocite{Still05}.

\section{XRT data analysis and results}
\label{xrt}
\indent

The XRT observations of the GRB 050318 field started on 2005 March 18 at 
16:39:11 UT, about 54 minutes after the BAT trigger. The instrument 
was in Auto State and the relatively low count rate level detected on the CCD 
automatically configured the instrument to its most sensitive ``Photon Counting'' 
operational mode (\cite{Hill04}) at 16:39:21 UT. The field was then observed for 
11 consecutive orbits until 09:25:48 UT on 2005 March 19.

The XRT data were processed with the 
XRTDAS\footnote{http://swift.gsfc.nasa.gov/docs/swift/analysis/xrt$\_$swguide$\_$v1$\_$2.pdf} 
software package (v.~1.4.0) developed at the ASI Science Data Center. 
Calibrated and cleaned Level 2 event files were produced with the {\it xrtpipeline} task. 
Only time intervals with a CCD temperature below $-50 ^\circ$C and events with grades 0--12 were 
considered for the data analysis. The total exposure time of data in 
Photon Counting mode after screening is 23540 s.

The 0.2--10 keV image of the field centered on GRB 050318 was analyzed with the 
XIMAGE package (v.~4.3). A previously uncatalogued bright X--ray source is clearly 
visible within the BAT error circle with coordinates 
RA(J2000) = $03^{\mathrm h}18^{\mathrm m}51.1^{\mathrm s}$, 
Dec(J2000) = $-46^{\circ}23'44.7''$. This position, determined with the 
{\sc detect} routine of XIMAGE which uses a sliding-cell method, has a 90\% uncertainty of 6 arcsec 
including systematic errors which take into account the residual 
calibration uncertainties in the spacecraft attitude and telescope misalignment. The XRT 
coordinates of the image centroid are 2.6 arcmin from the BAT centroid position and 
1.1 arcsec from the centroid position of the optical counterpart. Thus 
X--ray and optical counterparts are fully consistent with being the same object.

%
   \begin{figure}[ht]
   \centering
   \includegraphics[width=8.6cm]{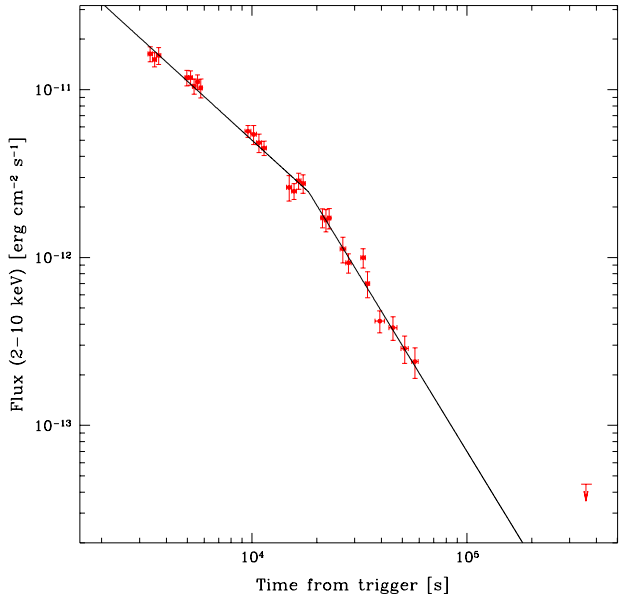}
      \caption{XRT 2--10 keV light curve of the afterglow of 
               GRB 050318. The solid line represents the broken power law 
               best fit model. 
              }
         \label{lc}
   \end{figure}

Events for the temporal analysis were selected within a circle of 20 pixel 
($\sim$47 arcsec) radius, which encloses about 90\% of the PSF at 1.5 
keV (\cite{moretti}), centered on the source position. The background was 
extracted from a nearby source-free circular region of 30 pixel radius. 
The 2--10 keV light curve, binned to that each bin contains at least 20 counts, 
is shown in Fig.~\ref{lc}. 
The X--ray afterglow of GRB 050318 is clearly fading. 
We first fitted the afterglow decay with a single power law model 
and found $\alpha = -1.44\pm 0.04$. However, this fit was clearly not acceptable, with
$\chi_r^2 = 3.05$ (25 dof). From the inspection of the 
residuals a clear steepening of the light curve with time was observed and 
a broken power law model with slopes $\alpha_1$,  $\alpha_2$ and break $t_{\rm{b}}$ 
was adopted. The model provided a very good fit with $\chi_r^2 = 0.80$ 
(23 dof) and best fit parameters $\alpha_1 = -1.17\pm 0.08$, $\alpha_2 = -2.10^{+0.22}_{-0.24}$ 
and $t_{\rm{b}} = 18345^{+~3362}_{-10508}~\mathrm{s}$.

The field of GRB 050318 was re-observed with the XRT on March 22 
starting at 01:09:50 UT until March 23 23:59:56 UT. In the Photon Counting 
mode exposure of 6394 s the X--ray afterglow was not detected. Its three-sigma 
upper-limit on the 2--10 keV flux, plotted in Fig.~\ref{lc}, is consistent with the 
extrapolation of the best fit model.

For the spectral analysis events were extracted from the same circular 
region used to produce the light curve. A further selection on XRT event 
grades 0--4 (i.e. single and double pixel events) was 
applied to the data. The spectrum was binned to ensure a minimum of 20 
counts per bin, and energy channels below 0.2 keV and above 10.0 keV were 
excluded, leaving a total of 1874 photons. The XRT average spectrum was 
fitted using the XSPEC package (v.~11.3.2, \cite{Arnaud96}). 
We adopted an absorbed power law model with energy index $\beta$ 
and the results of the spectral fit are shown 
in Table \ref{bestfit}. This model fits the data well 
($\chi_r^2=0.96$, 75 dof), with some evidence of a hydrogen-equivalent column density 
higher than the Galactic value ($N_{\rm H} = 5.2^{+1.3}_{-1.1}\times 10^{20}$~cm$^{-2}$ 
versus $N_{\rm H}^{\rm G} = (2.8\pm 1.0)\times 10^{20}~\mathrm{cm}^{-2}$, \cite{dickey}, 
\cite{Elvis94}). The 
average 2--10 keV flux corrected for absorption is $1.65 \times 10^{-12}$ 
\mbox{erg cm$^{-2}$ s$^{-1}$}. 
We also froze $N_{\rm H}$ to the Galactic value 
$N_{\rm H}^{\rm G}$ leaving the $N_{\rm H}^z$ redshifted to the rest frame of the GRB host 
($z=1.44$) free to vary ({\sc zwabs} model in {\sc XSPEC}). We found 
$N_{\rm H}^z = 1.3^{+0.7}_{-0.6}\times10^{21}$~cm$^{-2}$ as best fit with
$\beta$ unchanged with respect to the value reported in Table~\ref{bestfit}. 

\begin{table}[h]
\caption{Results of spectral fits of the X--ray afterglow of GRB 050318}
\label{bestfit}     
\centering                  
\begin{tabular}{c c c c}    
\hline\hline                
time interval & $\beta$ & $N_\mathrm{H}$ (cm$^{-2}$) & $\chi_r^2$ (d.o.f.) \\  
\hline                   
all      & $-1.09^{+0.09}_{-0.09}$ & 
  $(5.2^{+1.3}_{-1.1})\times10^{20}$ & 0.96 (75) \\ 
$t<t_{\rm b}$  & $-1.09^{+0.10}_{-0.11}$ & 
  $(5.1^{+1.5}_{-1.2})\times10^{20}$ & 1.22 (57) \\ 
$t>t_{\rm b}$  & $-1.06^{+0.24}_{-0.27}$ & 
  $(3.2^{+4.7}_{-3.2})\times10^{20}$ & 0.48 (20) \\ 
\hline                      
\end{tabular}
\end{table}

The spectral analysis was performed also in shorter time intervals to 
study the possible evolution of the spectrum with time. The 
observation was split in two segments, the first corresponding to the 
light curve before the break time $t_{\rm b}$, and the 
second one covering the remainder of the light curve. 
The results of the two spectral fits are shown in Table~\ref{bestfit}. 
No evidence for spectral variations of the X--ray afterglow is 
found.

\section{Discussion}
\label{discussion}
\indent

The results reported here, along with those of other GRBs detected by \swift\ 
(\cite{burrows05b}, \cite{Campana05}, \cite{Tagliaferri05}, \cite{cusumano}), 
clearly show that the study of the early afterglows and of the possible breaks in the
X--ray afterglow light curves is well within the  \swift\ possibilities. 
For GRB 050318 we have found a clear break in the X--ray afterglow light curve 
occurring at $t_{\rm b} \sim 18000$~s, with no evidence of spectral change (see Table 1). 
The slope of the light curve before the temporal break is $\alpha_1$$\sim$$-1.2$,
while that measured after $t_{\rm b}$ is $\alpha_2$$\sim$$-2.1$.
The  light curve of the GRB050318 optical counterpart
obtained with UVOT (\cite{Still05}), although  determined only before the break
observed with XRT, shows a slope consistent with the X--ray afterglow light curve.

In the framework of the fireball model, this fact points to a forward shock with 
the circumburst material responsible for the observed X--ray and optical
emission. The origin of the break was investigated. From the derived
estimates of $\alpha$ and $\beta$ before and after the temporal break,
assuming a synchrotron emission (see below),
it results that the break is inconsistent with the transition
of the cooling frequency from above to below the X--ray energy band
either when the fireball expands in a wind (\cite{Chevalier99}) or in a constant 
density medium (\cite{Sari98}). The temporal break also resulted to be 
inconsistent with the transition of the fireball to a non relativistic phase
(\cite{Dai99}).
It is therefore very likely to interpret the temporal break 
as due to the effect of a collimated relativistic outflow, when its
bulk Lorentz factor $\gamma$ becomes lower than the inverse of 
the jet opening angle $\theta_{\rm jet}$ (e.g., \cite{Rhoads97}, \cite{Sari99}).
In this framework, the jet opening angle  can be determined  through the equation 
$\theta_{\rm jet} = 0.161 [t_{\rm b}/(1+z)]^{3/8}(n \eta/E_{\rm iso})^
{1/8}$ (e.g., Bloom et al.~2003\nocite{Bloom03}) where $\theta_{\rm jet}$ is in radians, 
$t_{\rm b}$ in days, $E_{\rm iso}$ in units 10$^{52}$ erg, the density $n$ of the circumburst 
medium  in cm$^{-3}$, and $\eta$ is the efficiency of conversion of the outflow kinetic 
energy in electromagnetic radiation. With $z=1.44$ and thus with the $t_{\rm b}$ and $E_{\rm iso}$ 
values above derived (see Section~\ref{bat}),
we find $\theta_{\rm jet} = 3.65^{+0.25}_{-0.78}$ deg, for $\eta = 0.2$ 
(\cite{Frail01}) and assuming the new canonical value of circumburst 
density $n = 10$~cm$^{-3}$,
as discussed by Bloom et al.~(2003). With this value of the 
jet opening angle, the inferred collimation-corrected radiated energy is 
$E_{\gamma} = 4.5^{+0.7}_{-2.0} \times 10^{49}$~erg.

While the value of $E_{\rm iso}$ is within the range of values found for the GRBs with
known $z$, the value of $E_{\gamma}$ is in the tail ($\approx 2\sigma$) of the distribution 
reported by either Ghirlanda et al.~(2004)\nocite{Ghirlanda04} or Friedman 
\& Bloom (2005)\nocite{Friedman05}. Taking into account the $E_{{\rm p,rest}}$ value 
above derived (see Section~\ref{bat}), it is possible to see that GRB 050318 is 
fully consistent
with the $E_{{\rm p,rest}}$ vs. $E_{\rm iso}$ relation found by Amati et al.~(2002), while
is inconsistent ($\approx 3\sigma$ distance) with 
the $E_{{\rm p,rest}}$ vs. $E_{\gamma}$ relation 
found by Ghirlanda et al.~(2004). Assuming that the 
 90\% confidence interval (95--152 keV) of $E_{{\rm p,rest}}$ (see Sect.~\ref{bat}) 
is not strongly affected by systematics, the only viable way to make this event consistent
with the Ghirlanda relation, (a higher value of $\eta$ appears
unlikely) is to assume a higher circumburst density medium ($n \sim 100$~cm$^{-3}$). 
This would imply a higher $\theta_{\rm jet} (4.9^{+0.3}_{-1.0}$ deg) and 
$E_{\gamma}$ ($\sim 7.9^{+1.3}_{-3.5} \times 10^{49}$~erg).

   \begin{figure}
    \centering
    \includegraphics[width=8.2cm]{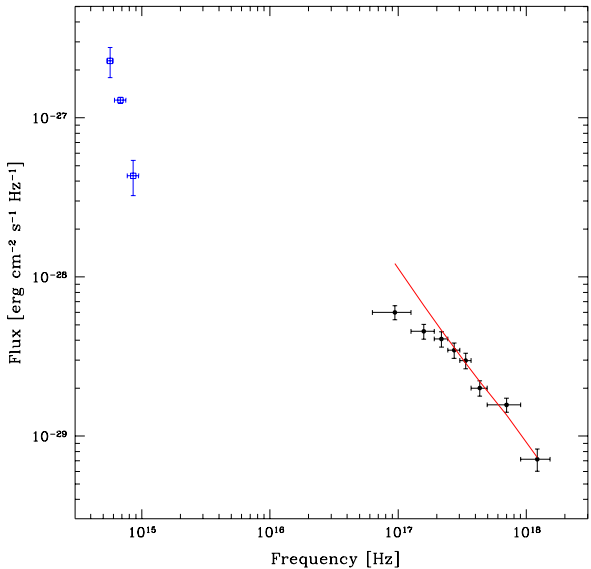}
      \caption{UVOT (V, B and U bands, open squares, Still et al.~(2005)) and XRT 
               (filled circles) multiwavelength 
               energy spectrum of the afterglow of GRB 050318 at epoch T+4061s after the trigger.
               The solid line is the spectral best fit model to the X--ray data 
               corrected for absorption.
              }
         \label{f:spectrum}
   \end{figure}
 
The X--ray afterglow spectrum is consistent with a synchrotron emission model 
from a spreading jet, confirming the given  interpretation of the temporal break
in the X--ray light curve.  Indeed, following Sari et al.~(1999), 
we find that, assuming a constant density medium and an adiabatic expansion 
of the fireball, in the slow cooling  regime, in which the peak frequency 
$\nu_{\rm m}$ is lower than the cooling frequency
$\nu_{\rm c}$, the slopes of the X--ray spectrum and light curve  are mutually consistent.
With these assumptions, the following closure
relations should be satisfied: for $t<t_{\rm b}$, $\alpha_1 = 3 \beta_1/2$ for $\nu<\nu_{\rm c}$, 
while $\alpha_1 =3\beta_1/2 + 1/2$ for 
$\nu>\nu_{\rm c}$; for  $t>t_{\rm b}$, $\alpha_2 = 2 \beta_2 - 1$ 
for $\nu<\nu_{\rm c}$, while $\alpha_2 = 2 \beta_2$ for $\nu>\nu_{\rm c}$. In our case, given
 that we do not find significant spectral evolution (see Table 1), we assume
$\beta_1 = \beta_2 = \beta = -1.09 \pm 0.09$. From the value derived for the
temporal index after the break ($-2.10^{+0.22}_{-0.24}$),  we can state
that $\nu<\nu_{\rm c}$ can be excluded (expected value of $\alpha_2 = -3.18 \pm 0.18$)
 while the case $\nu>\nu_{\rm c}$ is 
fully consistent with the data (expected $\alpha_2 = -2.18 \pm 0.18$). 
With $\nu>\nu_{\rm c}$, for $t>t_{\rm b}$,
the power-law index $p$ of the electron energy distribution ($N(E)\propto E^{-p}$),
that gives rise to the synchrotron photons, is expected to be coincident with
$-\alpha_2$ ($p \sim 2.1$), and thus in the range of values found for all the \sax\ GRBs 
(\cite{Frontera03}). With this value of $p$ we expect a
temporal index before the temporal break ($\alpha_1 = -3/4 p + 1/2 = -1.08 \pm 0.17$) 
and an energy index ($\beta = -p/2 = -1.05 \pm 0.11$),
both consistent with the  measured values ($\alpha_1 = -1.17\pm 0.08$, 
$\beta = -1.09 \pm 0.09$). We notice that, on the basis of the X--ray spectrum alone,
we cannot distinguish between a wind and a constant density medium: both media require
the same closure relations for $\nu >\nu_{\rm c}$ (\cite{Chevalier99}).

   \begin{figure}[t]
   \centering
    \includegraphics[width=8cm]{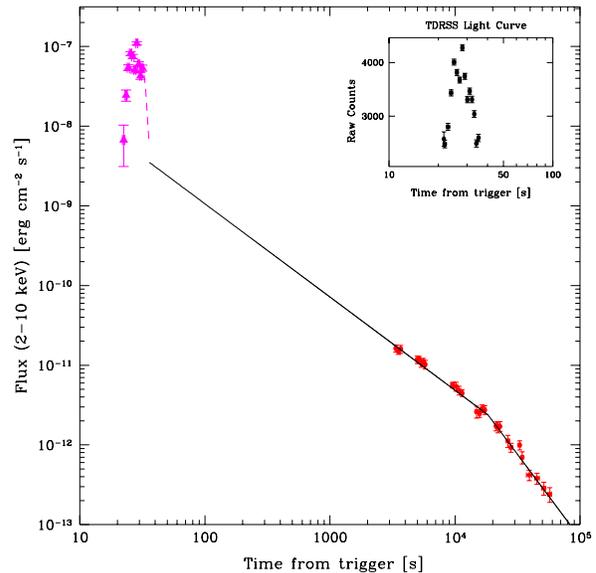}
     \caption{XRT 2--10 keV light curve (circles) of the afterglow of GRB 050318 compared 
              with the BAT light curve (triangles) of the second and third peaks. 
              The solid line is the best fit model to the XRT light curve. 
              The dashed line is an estimation of the last 2 
              seconds of the BAT light curve, during which event data were not 
	      recorded (\cite{krimm05b}), derived using the decay slope observed in the TDRSS data.
              Upper-right box: TDRSS 25--50 keV BAT light curve.
             }
         \label{lc2}
   \end{figure}

In the optical band the observational scenario is more 
complex. The measured temporal index  in the ${\rm U+V}$ band  ($\alpha_1^{\rm U+V} = 
-0.94 \pm 0.17$, Still et al.~2005\nocite{Still05}) is consistent with a synchrotron emission
model from a spreading jet.
In Fig.~\ref{f:spectrum} the multiwavelength spectrum of the afterglow is plotted. 
The UVOT optical points are taken from Still et al.~(2005) and refer to an epoch of 4061 seconds 
after the trigger. Accordingly, the XRT data have been selected in the same time interval of the 
UVOT exposures, i.e. between 3180 and 5822 seconds after the trigger (\cite{Still05}). As can be 
seen, it appears that the optical spectrum from V to U is much steeper than the X--ray spectrum 
(energy index $\beta = -4.9 \pm 0.5$, Still et al. 2005). The issue has been discussed by 
Still et al. (2005), finding that the XRT and UVOT
spectral data could be reproduced assuming either a gas and dust complex redshifted by $z = 2.8
\pm 0.3$ or dust extinction by more absorbing complexes at more moderate redshifts, like 
those discovered by Berger \& Mulchaey (2005). The above considerations on the $E_{\rm p}$ 
vs. $E_{\rm iso}$ and $E_{\rm p}$ vs. $E_\gamma$ relations remain unchanged even assuming $z =2.8$.

Finally, we focus on the connection between prompt and afterglow emission. In Fig.~\ref{lc2} the 
2--10 keV X--ray light curve of the afterglow emission from GRB 050318 is compared 
with the 25--50 keV prompt emission profile  measured with BAT, rescaled  to the 2--10 keV 
band using the BAT spectrum (see Section~\ref{bat}).
The back extrapolation of the afterglow light curve to the time of the prompt emission is 
somewhat below, but marginally consistent, with the end of the prompt light curve. However, 
uncertainties in the conversion factor from the 25--50 keV to the 2--10 keV energy band and the 
event data loss that affected the last $\sim$2 seconds of the burst (see Fig.~\ref{lc2}) do not 
allow us a more detailed investigation.
   
\section{Conclusions}
\label{conclusions}
\indent

On the basis of the \swift\ observations, GRB 050318 appears to be a 
classical GRB, with observational properties consistent with a collimated fireball
model, with a cone angle of 3--5~deg and a radiated energy $E_\gamma$ of
(4--8)$\times 10^{49}$ erg against an isotropic-equivalent energy $E_{\rm iso}$ of 
$\sim 2.2 \times 10^{52}$~erg, assuming a GRB redshift derived from
the farthest absorbing system discovered in the optical band ($z = 1.44$).
A search of the GRB host galaxy and its redshift should be crucial to confirm
these results and/or the inference of a $z= 2.8$ from the UVOT data.
In both cases the event fully satisfies the Amati relation and marginally the Ghirlanda
relation. The X--ray afterglow light curve and spectrum are
consistent with a synchrotron emission model during a slow cooling regime
with cooling frequency below the X--ray band.

\begin{acknowledgements}
We thank F. Tamburelli and B. Saija for their work on the XRT data reduction software.
This work is supported in Italy from ASI on contract number I/R/039/04 and through funding 
of the ASI Science Data Center, at Penn State by NASA contract NAS5-00136 and at the 
University of Leicester by the Particle Physics and Astronomy Research Council on grant 
numbers PPA/G/S/00524 and PPA/Z/S/2003/00507.
\end{acknowledgements}


\begin{thebibliography}{}

   \bibitem[Amati et al.~2002]{Amati02}
	Amati, L.. Frontera, F., Tavani, M. 2002, A\&A, 390, 81

   \bibitem[Arnaud 1996]{Arnaud96}
        Arnaud, K.A., 1996, Astronomical Data Analysis Software and Systems V, eds. Jacoby G. 
        and Barnes J., p17, ASP Conf. Series vol. 101

   \bibitem[Barthelmy et al.~2005]{barthelmy} 
       Barthelmy, S.D., Barbier, L.M., Cummings, J.R., et al., 2005, Space Science Rev., 120, 
       in press (astro-ph/0507410)

   \bibitem[Beardmore et al.~2005]{beardmore} 
       Beardmore, A.P., Page, K.L., Mangano, V., et al., 2005, GCN, 3133

   \bibitem[Berger \& Mulchaey 2005]{berger} 
       Berger, E., Mulchaey J., 2005, GCN, 3122

   \bibitem[Bloom et al.~2003]{Bloom03}
	Bloom, J.S., Frail, D.A., \& Kulkarni, S.R. 2003, ApJ, 594, 674

   \bibitem[Burrows et al.~2005a]{burrows05a} 
       Burrows, D.N., Hill, J.E., Nousek, J.A., et al., 2005a, Space Science Rev., 120, 
       in press (astro-ph/0508071)

   \bibitem[Burrows et al.~2005b]{burrows05b} 
       Burrows, D.N., Hill, J.E., Chincarini, G., et al., 2005b, 
       ApJ, 622, L85

   \bibitem[Campana et al.~2005]{Campana05} 
       Campana, S., Antonelli, L.A., Chincarini, G., et al., 2005, 
       ApJ, 625, L23

   \bibitem[Chevalier \& Li 1999]{Chevalier99}
   Chevalier, R.A. \& Li, Z.Y., 1999, ApJ, 520, L29

   \bibitem[Cusumano et al.~2005]{cusumano} 
       Cusumano, G., Mangano, V., Abbey, A.F., et al., 2005, 
       ApJ, submitted

   \bibitem[Dai \& Lu 1999]{Dai99}
	Dai, Z.G. \& Lu, T., 1999, ApJ, 519, L155

   \bibitem[De Pasquale et al.~2005]{depasquale} 
       De Pasquale, M., Boyd, P., Holland, S.T., et al., 2005, GCN, 3123

   \bibitem[Dickey \& Lockman 1990]{dickey} 
       Dickey, J.M. \& Lockman., F.J., 1990, ARA\&A, 28, 215
   
   \bibitem[Elvis et al.~1994]{Elvis94}
	Elvis, M., Lockman, F.J. and Fassnacht, C., 1994, ApJS, 95, 413

   \bibitem[Frail et al.~2001]{Frail01}
	Frail, D.A., Kulkarni, S.R., Sari, R. et al., 2001, A\&A, 562, L55

   \bibitem[Friedmann \& Bloom]{Friedman05}
	Friedmann, A.S. \& Bloom, J.S., 2005, ApJ, 627, 1
 
   \bibitem[Frontera et al.~2000]{Frontera00}
	Frontera, F., Amati, L., Costa, E. et al., 2000, ApJS, 127, 59

   \bibitem[Frontera 2003]{Frontera03}
	Frontera, F. 2003, in {\it Supernovae and Gamma--Ray Bursters},
	ed. K. Weiler (Springer, Berlin), Lecture Notes in Physics, 598, 317

   \bibitem[Gehrels et al.~2004]{gehrels} 
       Gehrels, N., Chincarini, G., Giommi, P., et al., 2004, ApJ, 611, 1005

   \bibitem[Ghirlanda et al.~2004]{Ghirlanda04}
	Ghirlanda, G., Ghisellini, G., \& Lazzati, D. 2004, ApJ, 616, 331

   \bibitem[Hill et al.~2004]{Hill04}
        Hill, J.E., Burrows, D.N., Nousek, J.A., et al., 2004, Proceedings of SPIE, Vol. 5165, 217   
 
   \bibitem[Krimm et al.~2005a]{krimm05a} 
       Krimm, H., Barthelmy, S., Barbier, L., et al., 2005a, GCN, 3111

   \bibitem[Krimm et al.~2005b]{krimm05b} 
       Krimm, H., Barthelmy, S., Barbier, L., et al., 2005b, GCN, 3134

   \bibitem[McGowan et al.~2005]{mcgowan} 
       McGowan, K., De Pasquale, M., Boyd, P., et al., 2005, GCN, 3115

   \bibitem[Maiorano et al.~2005]{Maiorano05}
	Maiorano, E., Masetti, N., Palazzi, E., et al., 2005, A\&A, in press (astro-ph/0504602)

   \bibitem[Moretti et al.~2004]{moretti} 
       Moretti, A., Campana, S., Tagliaferri, G., et al., 2004, Proceedings of SPIE, Vol. 5165, 232

   \bibitem[Mulchaey \& Berger 2005]{mulchaey} 
       Mulchaey, J., Berger, E., 2005, GCN, 3114

   \bibitem[Nousek et al.~2005]{nousek} 
       Nousek, J.A., Morris, D.C., Burrows, D.N., et al., 2005, GCN, 3113

   \bibitem[Page et al.~2005]{page05} 
       Page, K.L., Rol, E., Levan, A.J., et al., 2005, MNRAS, in press (astro-ph/0508011)

   \bibitem[Pian et al.~2001]{Pian01}
	Pian, E., Soffitta, P., Alessi, A., et al., 2001, A\&A, 372, 456

   \bibitem[Rhoads 1997]{Rhoads97}
	Rhoads, J. 1997, ApJ, 487, L1

   \bibitem[Roming et al.~2005]{roming} 
       Roming, P.W.A., Kennedy, T.E., Mason, K.O., et al., 2005, Space Science Rev., 120, 
       in press (astro-ph/0507413)
   
   \bibitem[Sari et al.~1998]{Sari98}
	Sari, R., Piran, T., \& Narayan, N., 1998, ApJ, 497, L17
 
   \bibitem[Sari et al.~1999]{Sari99}
	Sari, R., Piran, T. \& Halpern, J.P. 1999, ApJ, 519, L17
 
   \bibitem[Still et al.~2005]{Still05} 
       Still, M., Roming, P.W.A., Mason, K.O., et al., 2005, ApJ, in press (astro-ph/0509060)

   \bibitem[Tagliaferri et al.~2005]{Tagliaferri05} 
       Tagliaferri, G., Goad, M.R., Chincarini, G., et al., 2005, Nature, 436, 985

   \bibitem[in't Zand et al.~2001]{Zand01}
       in't Zand, J.J.M., Kuiper, L., Amati, L. et al., 2001, ApJ, 559, 710

\end{thebibliography}
\end{document}